\RequirePackage{ifpdf}
\ifpdf 
\documentclass[pdftex]{sigma}
\else
\documentclass{sigma}
\fi

\def\Cbbd{\mathbb{C}}

\def\Acal{\mathcal{A}}
\def\Bcal{\mathcal{B}}

\def\Gcal{\mathcal{G}}

\def\Lcal{\mathcal{L}}
\def\Mcal{\mathcal{M}}
\def\Pcal{\mathcal{P}}
\def\Tcal{\mathcal{T}}

\newcommand{\rd}{\mathrm{d}}            
\newcommand{\re}{\mathrm{e}}            
\def\a{\alpha}
\def\b{\beta}

\def\G{\Gamma}
\def\d{\delta}
\def\eps{\varepsilon}
\def\la{\lambda}
\def\La{\Lambda}
\def\s{\sigma}
\def\phi{\varphi}

\def\om{\omega}
\def\Om{\Omega}
\def\BT{B\"{a}cklund transformation}
\def\tr{\mathop{\hbox{\rm tr}}\nolimits}

\def\dd{\partial}
\def\Ref#1{(\ref{#1})}
\def\tilde{\widetilde}

\def\tphi{\tilde{\phi}}

\def\?{(?)\marginpar{|?}}

\def\beq{\begin{equation}}
\def\eeq{\end{equation}}
\def\be{\begin{equation*}}
\def\ee{\end{equation*}}

\begin{document}

\numberwithin{equation}{section}

\allowdisplaybreaks

\renewcommand{\PaperNumber}{080}

\FirstPageHeading

\renewcommand{\thefootnote}{$\star$}

\ShortArticleName{B\"{a}cklund Transformation for the BC-Type Toda Lattice}

\ArticleName{B\"{a}cklund Transformation \\ for the BC-Type Toda Lattice\footnote{This paper is a
contribution to the Vadim Kuznetsov Memorial Issue `Integrable
Systems and Related Topics'. The full collection is available at
\href{http://www.emis.de/journals/SIGMA/kuznetsov.html}{http://www.emis.de/journals/SIGMA/kuznetsov.html}}}

\Author{Vadim KUZNETSOV~$^\dag$ and Evgeny SKLYANIN~$^\ddag$}

\AuthorNameForHeading{V. Kuznetsov and E. Sklyanin}

\Address{$^\dag$~Deceased}
\Address{$^\ddag$~Department of Mathematics, University of York,
York YO10 5DD, UK}
\EmailD{\href{mailto:eks2@york.ac.uk}{eks2@york.ac.uk}}

\ArticleDates{Received July 13, 2007; Published online July 25, 2007}

\Abstract{We study an integrable case of $n$-particle Toda lattice:
open chain with boundary terms containing 4 parameters.
For this model we construct a \BT\ and prove its basic properties:
canonicity, commutativity and spectrality.
The \BT\ can be also viewed as a discretized time dynamics.
Two Lax matrices are used: of order 2 and of order $2n+2$, which are
mutually dual, sharing the same spectral curve.}


\Classification{70H06}

\section{Introduction}\label{intro}

In the present paper we study the Hamiltonian system of $n$ one-dimensional
particles with coordinates $x_j$ and canonical momenta $X_j$,
$j=1,\ldots,n$:
\beq
  \{X_j,X_k\}=\{x_j,x_k\}=0,\qquad \{X_j,x_k\}=\d_{jk},
\label{pbXx}
\eeq
characterized by the Hamiltonian
\beq
   H=\sum_{j=1}^n \frac12 X_j^2+\sum_{j=1}^{n-1}\re^{x_{j+1}-x_j}
     +\a_1\re^{x_1}+\frac12\b_1\re^{2x_1}
     +\a_n\re^{-x_n}+\frac12\b_n\re^{-2x_n}
\label{hamiltonian}
\eeq
containing 4 arbitrary parameters: $\a_1$, $\b_1$, $\a_n$, $\b_n$.

The model
was missing from the early lists of integrable cases
of the Toda lattice \cite{Bog76,AvM82}
based on Dynkin diagrams for simple af\/f\/ine Lie algebras.
Its integrability was proved f\/irst in \cite{Skl87,Skl88,Ino}.
As for the more recent classif\/ications, in \cite{KT89} the model is enlisted as the case (i).
In \cite{RS94,RS2003} particular cases of the Hamiltonian \Ref{hamiltonian}
are assigned to the $C_n^{(1)}$ case with `Morse terms'.
For brevity, we refer to the model as `BC-Toda lattice' emphasising the fact that
each boundary term is a linear combination of the term $\sim\a$ corresponding to
the root system $B$ and of the term $\sim\b$ corresponding to
the root system $C$, see \cite{Bog76,AvM82,RS94,RS2003}.

In section \ref{integrability} we review brief\/ly the known facts about the integrability
of the model using the approach developed in \cite{Skl87,Skl88} and
based on the Lax matrix $L(u)$ of order 2 and the corresponding quadratic $r$-matrix
algebra. In particular, we construct explicitly a generating
function of the complete set of commuting Hamiltonians $H_j$ $(j=1,\ldots, n$)
which includes the physical Hamiltonian $H$ \Ref{hamiltonian}.

In Section \ref{BTdef} we describe the main result of our paper:
construction of a \BT\ (BT) for our model
as a one-parametric family of maps $\Bcal_\la:(Xx)\mapsto(Yy)$
from the variables $(Xx)$ to the variables $(Yy)$.
We construct the BT choosing an appropriate gauge (or Darboux) transformation
of the local Lax matrices. In Section \ref{BTprop},
adopting the Hamiltonian point of view
developed in \cite{KS98,Skl99}, we prove the basic properties of the BT:
\begin{enumerate}\itemsep=0pt
\item Preservation of the commuting Hamiltonians $\Bcal_\la: H_j(X,x)\mapsto H_j(Y,y)$.
\item Canonicity: preservation of the Poisson bracket \Ref{pbXx}.
\item Commutativity: $\Bcal_{\la_1}\circ\Bcal_{\la_2}=\Bcal_{\la_2}\circ\Bcal_{\la_1}$.
\item Spectrality: the fact that the graph of the BT is a Lagrangian manifold
on which the 2-form
\beq
    \Omega\equiv\sum_{j=1}^n\bigl(\rd X_j\wedge\rd x_j-\rd Y_j\wedge\rd y_j\bigr)-\rd\ln\La\wedge\rd\la
\eeq
vanishes. Here $\La$ is an eigenvalue of the matrix $L(\la)$.
In other words,
the parameter $\la$ of the BT and its exponentiated canonical conjugate $\La$
lie on the spectral curve of $L(u)$:
\beq
    \det\bigl(\La-L(\la)\bigr)=0.
\label{eq:spectrality}
\eeq
\end{enumerate}

We also prove the following expansion of $\Bcal_\la$ in $\la^{-1}$
\beq
    \Bcal_\la:f\mapsto f-2\la^{-1}\{H,f\}+O(\la^{-2}), \qquad
    \la\rightarrow\infty.
\label{expand_B}
\eeq
which allows to interpret the BT as a discrete time dynamics
approximating the continuous-time dynamics generated by the Hamiltonian~\Ref{hamiltonian}.

In Section \ref{dualL} we construct for our system an alternative Lax matrix
$\Lcal(v)$. The new Lax matrix of order $2n+2$ is dual to the matrix $L(u)$
of order 2 in the sense that they share the same spectral curve with the parameters
$u$ and $v$ having been swapped:
\beq
    \det\bigl(v-L(u)\bigr)=(-1)^{n+1}v\det\bigl(u-\Lcal(v)\bigr).
\label{dual_curve}
\eeq

In the same section
we provide an interpretation of the BT
in terms of the `big' Lax mat\-rix~$\Lcal(v)$
and establish a remarkable factorization formula for $\la^2-\Lcal^2(v)$.

The concluding Section \ref{discussion}
contains a summary and a discussion.
All the technical proofs and tedious calculations are removed to the Appendices.

\section{Integrability of the model}\label{integrability}

In demonstrating the integrability of the model we follow the
approach to the integrable chains with boundary conditions
developed in \cite{Skl87,Skl88} and use the notation
of \cite{KS98,Skl99}.

The Lax matrix $L(u)$ for the BC-Toda lattice is constructed
as the product
\beq
  L(u)=K_-(u)T^t(-u)K_+(u)T(u)
\label{defL}
\eeq
of the following matrices ($T^t$
stands for the matrix transposition).

The monodromy matrix $T(u)$
is itself the product
\beq
  T(u)=\ell_n(u)\cdots\ell_1(u)
\label{defT}
\eeq
of the local Lax matrices
\beq
  \ell_j(u)\equiv\ell(u;X_j,x_j)
            =\begin{pmatrix}
             u+X_j & -\re^{x_j} \\
             \re^{-x_j} & 0
             \end{pmatrix},
\label{defell}
\eeq
each containing only the variables $X_j$, $x_j$ describing
a single particle. Note that $\tr T(u)$ is the generating
function for the Hamiltonians of the periodic Toda lattice.

The matrices $K_\pm(u)$ containing
the information about the boundary interactions
are def\/ined as \cite{Skl87,Skl88}
\beq
  K_-(u)=\begin{pmatrix}
             u & -\a_1 \\
             \a_1 & \b_1 u
               \end{pmatrix}, \qquad
  K_+(u)=\begin{pmatrix}
             u & -\a_n \\
             \a_n & \b_n u
             \end{pmatrix}.
\label{defKpm}
\eeq

The signif\/icance of the Lax matrix $L(u)$ is that
its spectrum is invariant under the dynamics generated by
the Hamiltonian \Ref{hamiltonian}, the corresponding equations
of motion $\rd G/\rd t\equiv\dot{G}=\{H,G\}$
for an observable $G$ being
\beq\label{eq:dynamics-x}
    \dot{x}_j=X_j, \qquad j=1,\ldots,n
\eeq
and
\begin{subequations}\label{eq:dynamics-X}
\begin{gather}
   \dot{X}_j=\re^{x_{j+1}-x_j}-\re^{x_j-x_{j-1}}, \qquad j=2,\ldots,n-1, \\
   \dot{X}_1=\re^{x_2-x_1}-\a_1\re^{x_1}-\b_1\re^{2x_1}, \\
   \dot{X}_n=-\re^{x_n-x_{n-1}}+\a_n\re^{-x_n}+\b_n\re^{-2x_n}.
\end{gather}
\end{subequations}

To prove the invariance of the spectrum of $L(u)$
we introduce the matrices $A_j(u)$
\begin{gather}
    A_j(u)=\begin{pmatrix}
             -u & \re^{x_j} \\
             -\re^{-x_{j-1}} & 0
           \end{pmatrix}, \qquad j=2,\ldots,n-1,
\label{defAj}
\\
    A_1(u)=\begin{pmatrix}
             -u & \re^{x_1} \\
             -\a_1-\b_1\re^{x_1} & 0
           \end{pmatrix}, \qquad
    A_{n+1}(u)=\begin{pmatrix}
            -u & \a_n+\b_n\re^{-x_n} \\
            -\re^{-x_n} & 0
           \end{pmatrix},
\label{defA1n+1}
\end{gather}
which satisfy the easily verif\/ied identities
\setcounter{equation}{9}
\begin{subequations}\label{dotK}
\begin{gather}
    \dot{\ell}_j=A_{j+1}\ell_j-\ell_j A_j, \qquad j=1,\ldots,n,
\label{elldot}\tag{2.9}
\\
  -\dot{K}_+=0=K_+A_{n+1}(u)+A^t_{n+1}(-u)K_+, \\
  \dot{K}_-=0=A_1(u)K_-+K_-A^t_1(-u).
\end{gather}
\end{subequations}

From \Ref{defT} and \Ref{elldot} it follows immediately that
\beq
    \dot{T}(u)=A_{n+1}(u)T(u)-T(u)A_1(u).
\eeq

Then, using \Ref{defL} and \Ref{dotK}, we obtain the equality
\beq
   \dot{L}(u)=\bigl[A_1(u),L(u)\bigr]
\label{Ldot}
\eeq
implying that the spectrum of $L(u)$ is preserved
by the dynamics.

There are only two spectral invariants of a $2\times2$ matrix:
the trace and the determinant. From~\Ref{defell} it follows
that $\det\ell(u)=1$ and, respectively, $\det T(u)=1$,
so, by \Ref{defL}, the determinant of $L(u)$
\beq
    d(u)\equiv\det L(u)=\det K_-(u)\det K_+(u)=(\a_1^2+\b_1u^2)(\a_n^2+\b_nu^2)
\label{detL}
\eeq
contains no dynamical variables $Xx$.
The trace
\beq
  t(u)\equiv\tr L(u)=\tr K_-(u)T^t(-u)K_+(u)T(u),
\eeq
however, does contain dynamical variables and therefore
can be used as a generating function of the integrals of motion,
which can be chosen as the coef\/f\/icients of the polynomial $t(u)$
of degree $2n+2$ in $u$.
Note that $t(-u)=t(u)$ due to the symmetry
\beq
  K_\pm^t(-u)=-K_\pm(u).
\eeq

The leading coef\/f\/icient of $t(u)$ at $u^{2n+2}$ is a constant
$(-1)^n$. Same is true for its free term
\beq
   t(0)=\tr K_+(0)K_-(0)=-2\a_n\a_1
\eeq
due to the identity
\beq
  MK_\pm(0)M^t=\det M\cdot K_\pm(0),
\eeq
which holds for any matrix $M$.

We are left then with $n$ nontrivial coef\/f\/icients $H_j$
\beq
    t(u)=(-1)^nu^{2n+2}-2\a_n\a_1+\sum_{j=1}^n H_j u^{2j}
\eeq
which are integrals of motion $\dot{H}_j=0$
since $\dot{t}(u)=0$ due to \Ref{Ldot}.

The conserved quantities $H_j$ are obviously polynomial in $X$, $\re^{\pm x}$.
Their independence can easily be established by setting $\re^{\pm x}=0$
in \Ref{defell}
and analysing the resulting polynomials in $X$.
It is also easy to verify that the physical Hamiltonian \Ref{hamiltonian}
is expressed as
\beq
    H=\frac{(-1)^{n+1}}{2}\,H_n.
\eeq

The quantities $H_j$ are also in involution
\beq
    \{H_j,H_k\}=0
\label{comH}
\eeq
with respect to the Poisson bracket \Ref{pbXx}.
Together with the independence of $H_j$,
it constitutes the Liouville integrability of our system.

The commutativity \Ref{comH} of $H_j$ or, equivalently,
of $t(u)$
\beq
    \{t(u_1),t(u_2)\}=0
\label{com_t}
\eeq
is proved in the standard way using the $r$-matrix technique \cite{Skl87,Skl88}.

Let $\boldsymbol{1}$ be the unit matrix of order 2 and for any matrix $L$
def\/ine
\beq
    \overset{1}{L}\equiv L\otimes\boldsymbol{1}, \qquad
    \overset{2}{L}\equiv \boldsymbol{1} \otimes L.
\eeq

We have then the quadratic Poisson brackets \cite{Skl99,FTbook}
\beq
  \{\overset{1}\ell(u_1),\overset{2}\ell(u_2)\}=
  [r(u_1-u_2),\overset{1}\ell(u_1)\overset{2}\ell(u_2)],
\label{rll}
\eeq
and, as a consequence,
\beq
  \{\overset{1}T(u_1),\overset{2}T(u_2)\}=
  [r(u_1-u_2),\overset{1}T(u_1)\overset{2}T(u_2)],
\eeq
with the  $r$-matrix{\samepage
\beq\label{def-r}
  r(u)=\frac{\Pcal}{u},
\eeq
where $\Pcal$ is the permutation matrix $\Pcal a\otimes b=b\otimes a$.}

Let
\beq
  \tilde r(u)=r^{t_1}(u)=r^{t_2}(u),
\eeq
$t_1$ and $t_2$ being, respectively, transposition with respect to
the f\/irst and second component of the tensor product $\Cbbd^2\otimes\Cbbd^2$.

Then for both $\Tcal(u)=T(u)K_-(u)T^t(-u)$ and $\Tcal(u)=T^t(-u)K_+(u)T(u)$
we obtain the same Poisson algebra \cite{Skl87,Skl88}
\begin{gather}
 \{\overset{1}\Tcal(u_1),\overset{2}\Tcal(u_2)\}=
  r(u_1-u_2)\overset{1}\Tcal(u_1)\overset{2}\Tcal(u_2)
  -\overset{1}\Tcal(u_1)\overset{2}\Tcal(u_2)r(u_1-u_2) \notag\\
\phantom{\{\overset{1}\Tcal(u_1),\overset{2}\Tcal(u_2)\}=}{} -\overset{1}\Tcal(u_1)\tilde r(u_1+u_2)\overset{2}\Tcal(u_2)
  +\overset{2}\Tcal(u_2)\tilde r(u_1+u_2)\overset{1}\Tcal(u_1),
\label{TT}
\end{gather}
which ensures the commutativity \Ref{com_t} of $t(u)$.

\section{Describing \BT}\label{BTdef}

In this section we shall construct a \BT\ (BT) for our model.
We shall stay in the framework of the Hamiltonian approach
proposed in \cite{KS98} and follow closely our previous treatment
of the periodic Toda lattice \cite{KS98,Skl99}, with the
necessary modif\/ications taking into account the boundary conditions.

We are looking thus for a one-parametric family of maps
$\Bcal_\la:(Xx)\mapsto(Yy)$
from the variables $(Xx)$ to the variables $(Yy)$
characterised
by the properties enlisted in
the Introduction: {\it Invariance of Hamiltonians,
Canonicity, Commutativity and Spectrality.}

The invariance of the commuting Hamiltonians $H_j$,
or of their generating polynomial $t(u)=\tr L(u)$
will be ensured if we f\/ind an invertible matrix
$M_1(u,\la)$ intertwining the matrices $L(u)$
depending on the variables $Xx$ and $Yy$:
\beq
    M_1(u,\la)L(u;Y,y)=L(u;X,x)M_1(u,\la).
\label{ML}
\eeq

To f\/ind $M_1(u,\la)$ let us look for a
gauge transformation
\beq
   M_{j+1}(u,\la)\ell(u;Y_j,y_j)=\ell(u;X_j,x_j)M_j(u,\la), \qquad
   j=1,\ldots,n,
\label{Mel}
\eeq
implying that $\det M_j$ does not depend on $j$.
From \Ref{Mel} and \Ref{defT} we obtain
\beq
    M_{n+1}(u,\la)T(u;Y,y)=T(u;X,x)M_1(u,\la).
\label{MT}
\eeq

Let $J$ be the the standard skew-symmetric matrix of order 2
\beq
    J=\begin{pmatrix}
      0 & 1 \\ -1 & 0
       \end{pmatrix}, \qquad
       J^t=-J, \qquad J^2=-\boldsymbol{1},
\eeq
and def\/ine the antipode $M^a$ as
\beq
    M^a\equiv -JMJ
\label{antipode}
\eeq
for any matrix $M$ of order 2. It is easy to see that
\beq
    M^tM^a=M^aM^t=\det M.
\label{MtMa}
\eeq

Transposing \Ref{MT} and using \Ref{MtMa} together with the the fact that
$\det M_j$ is independent of $j$ we obtain the relation
\beq
    T^t(-u;X,x)M_{n+1}^a(-u,\la)=M_1^a(-u,\la)T^t(-u;Y,y).
\label{M1Tt}
\eeq

We shall be able to obtain \Ref{ML}
if we impose two additional relations
\begin{subequations}\label{MK_boundary}
\begin{gather}
  K_-(u)M^a_1(-u,\la)=M_1(u,\la)K_-(u),
\label{M1KM1} \\
  K_+(u)M_{n+1}(u,\la)=M^a_{n+1}(-u,\la)K_+(u).
\label{MnKMn}
\end{gather}
\end{subequations}

Then, starting with the right-hand side $L(u;X,x)M_1(u,\la)$
of \Ref{ML} and using \Ref{defL} and \Ref{MT}
we obtain
\begin{gather}
    L(u;X,x)M_1(u,\la)=K_-(u)T^t(-u;X,x)K_+(u)T(u;X,x)M_1(u,\la) \notag\\
\phantom{L(u;X,x)M_1(u,\la)}{} =K_-(u)T^t(-u;X,x)K_+(u)M_{n+1}(u,\la)T(u;Y,y)
\end{gather}

Using then \Ref{MnKMn}
to move $M_{n+1}(u,\la)$ through $K_+(u)$,
then using \Ref{M1Tt} and f\/inally \Ref{M1KM1}
we get, step by step,
\begin{gather}
   L(u;X,x)M_1(u,\la)=K_-(u)T^t(-u;X,x)M_{n+1}^a(-u,\la)K_+(u)T(u;Y,y) \notag\\
 \phantom{L(u;X,x)M_1(u,\la)}{}=K_-(u)M_1^a(-u,\la)T^t(-u;Y,y)K_+(u)T(u;Y,y) \notag \\
\phantom{L(u;X,x)M_1(u,\la)}{}=M_1(u,\la)K_-(u)T^t(-u;Y,y)K_+(u)T(u;Y,y) \notag \\
\phantom{L(u;X,x)M_1(u,\la)}{}=M_1(u,\la)L(u;Y,y)
\end{gather}
arriving f\/inally at \Ref{ML}.

We have thus to f\/ind a set of matrices $M_j(u,\la)$, $j=1,\ldots,n+1$
compatible with the conditions~\Ref{Mel} and~\Ref{MK_boundary}.
A quick calculation shows that the so called DST-ansatz for $M_j$
used in \cite{KS98,Skl99} for the periodic Toda lattice
contradicts the conditions \Ref{MK_boundary}.

The philosophy advocated in \cite{Skl99} requires that the ansatz for the gauge matrix
$M_j(u)$ be chosen in the form of a Lax matrix satisfying the $r$-matrix
Poisson bracket \Ref{rll} with the same $r$-matrix \Ref{def-r} as the Lax operator
$\ell(u)$. It was shown in \cite{Skl99} that the so-called DST-ansatz
\beq
   M^{\text{DST}}_j(u,\la)=\begin{pmatrix}
             u-\la+s_jS_j & -s_j \\ S_j & -1
            \end{pmatrix}
\eeq
serves well for the periodic Toda case. The above ansatz is however not compatible
with the boundary conditions \Ref{MK_boundary} and we have to use a more complicated
ansatz for $M_j$ in the form of the Lax matrix for the
isotropic Heisenberg magnet (XXX-model):
\beq
  M_j(u,\la)=\begin{pmatrix} u-\la+s_jS_j & s_j^2S_j-2\la s_j \\
              S_j & -u-\la+s_jS_j
        \end{pmatrix}, \qquad
        \det M_j(u,\la)=\la^2-u^2.
\label{def_Mj}
\eeq

The same gauge transformation was used in \cite{DM05}
for constructing a $Q$-operator for the quantum XXX-magnet.

Substituting \Ref{def_Mj}
into \Ref{Mel} we obtain the relations
\begin{subequations}\label{XYSS}
\begin{gather}
  X_j=-\la+s_j^{-1}\re^{x_j}+s_{j+1}\re^{-x_j}, \\
  Y_j=\la-s_j^{-1}\re^{y_j}-s_{j+1}\re^{-y_j}, \\
  S_j=2\la s_j^{-1}-s_j^{-2}\re^{x_j}-s_j^{-2}\re^{y_j}, \\
  S_{j+1}=\re^{-x_j}+\re^{-y_j},\label{Sj1expand}
\end{gather}
\end{subequations}
for $j=1,\ldots,n$,
and from \Ref{MK_boundary}, respectively,
\beq
  S_1=\frac{2(\a_1+\b_1\la s_1)}{1+\b_1 s_1^2}, \qquad
  S_{n+1}=\frac{2(\la s_{n+1}-\a_n)}{\b_n+s_{n+1}^2}.\label{expandS1}
\eeq

Eliminating the variables $S_j$, we arrive to the equations
def\/ining the BT $(j=1,\ldots,n)$:
\begin{subequations}\label{XTbcklnd}
\begin{gather}
 X_j=-\la+s_j^{-1}\re^{x_j}+s_{j+1}\re^{-x_j}, \\
 Y_j=\la-s_j^{-1}\re^{y_j}-s_{j+1}\re^{-y_j}.
\end{gather}
\end{subequations}

The variables $s_j$, $j=1,\ldots,n+1$ in \Ref{XTbcklnd} are implicitly def\/ined as
functions of $x$, $y$ and $\la$ from the quadratic equations
\begin{subequations}\label{eq:def-s}
\begin{gather}
 (\re^{-x_{j-1}}+\re^{-y_{j-1}})s_j^2
 -2\la s_j+(\re^{x_j}+\re^{y_j})=0, \qquad j=2,\ldots,n \label{eq_sj}\\
 (2\a_1+\b_1 \re^{x_1}+\b_1 \re^{y_1})s_1^2
 -2\la s_1+(\re^{x_1}+\re^{y_1})=0, \label{eq_s1}\\
 (\re^{-x_n}+\re^{-y_n})s_{n+1}^2-2\la s_{n+1}
 +(2\a_n+\b_n \re^{-x_n}+\b_n \re^{-y_n})=0.\label{eq_sn+1}
\end{gather}
\end{subequations}

Like in the periodic case \cite{KS98,Skl99},
the BT map $\Bcal_\la:(Xx)\mapsto(Yy)$
is described implicitly by the equations \Ref{XTbcklnd}.
Unlike the periodic case, we have extra variables $s_j$.
It is more convenient not to express $s_j$ from equations
\Ref{eq:def-s} and to substitute them into \Ref{XTbcklnd}
but rather def\/ine the BT by the whole set of equations
\Ref{XTbcklnd} and \Ref{eq:def-s}.

Equations \Ref{XTbcklnd} and \Ref{eq:def-s} are algebraic equations and
therefore def\/ine $(Yy)$ as multivalued functions of $(Xx)$,
which is a common situation with integrable maps \cite{Ves91}.

In this paper, to avoid the complications of the real algebraic geometry
we allow all our variables to be complex.

\section{Properties of the \BT}\label{BTprop}

Having def\/ined the map $\Bcal_\la:(Xx)\mapsto(Yy)$ in the previous section,
we proceed to establish its properties from the list given in the Introduction.

\subsection{Preservation of Hamiltonians} The equality $H_j(X,x)=H_j(Y,y)$
$\forall\; \la$, or, equivalently, $t(u;X,x)=t(u;Y,y)$
holds by construction, being a direct consequence of \Ref{ML}.

\subsection{Canonicity}\label{Canonicity}
The canonicity of the BT means that the variables
$Y(X,x;\la)$ and $y(X,x;\la)$ have the same canonical Poisson brackets \Ref{pbXx}
as $(Xx)$. An equivalent formulation can be given in terms of symplectic spaces and
Lagrangian manifolds. Consider the $4n$-dimensional symplectic space $V_{4n}$
with coordinates $XxYy$ and symplectic 2-form
\beq
    \Om_{4n}\equiv\sum_{j=1}^n\bigl(\rd X_j\wedge\rd x_j-\rd Y_j\wedge\rd y_j\bigr).
\label{defOm4n}
\eeq

Equations \Ref{XTbcklnd} and \Ref{eq:def-s}
def\/ine a $2n$-dimensional submanifold $\G_{2n}\subset V_{4n}$
which can be considered as the graph $Y=Y(X,x;\la)$, $y=y(X,x;\la)$ of the BT
(the parameter $\la$ is assumed here to be a constant).
The canonicity of the BT is then equivalent to the fact
that the manifold $\G_{2n}$ is {\it Lagrangian},
meaning that: (a) it is {\it isotropic}, that is nullif\/ies the form $\Om_{4n}$
\beq
\left.\Om_{4n}\right|_{\G_{2n}}=0,
\eeq
and (b) it has maximal possible dimension for an isotropic manifold:
$\dim \G_{2n}=\frac12\dim V_{4n}$.

One way of proving the canonicity is to present explicitly
the generating function $\Phi_\la(y;x)$ of the canonical transformation,
such that
\beq
   X_j=\frac{\dd\Phi_\la}{\dd x_j}, \qquad
   Y_j=-\frac{\dd\Phi_\la}{\dd y_j}.
\label{XYPhi}
\eeq

The required function is given by the expression
\beq
    \Phi_\la(y;x)=\sum_{j=1}^n f_\la(y_j,s_{j+1};x_j,s_j)
    +\phi_\la^{(0)}(s_1)+\phi_\la^{(n+1)}(s_{n+1}),
\label{def_Phi}
\eeq
where
\begin{subequations}
\begin{gather}
   f_\la(y_j,s_{j+1};x_j,s_j)=\la(2\ln s_j-x_j-y_j)
        +s_j^{-1}(\re^{x_j}+\re^{y_j})-s_{j+1}(\re^{-x_j}+\re^{-y_j}), \\
   \phi_\la^{(0)}(s_1)=-\la\ln\bigl(1+\b_1s_1^2)
          -\frac{2\a_1}{\sqrt{\b_1}}\arctan\bigl(\sqrt{\b_1}s_1\bigr), \label{def_phi0}\\
   \phi_\la^{(n+1)}(s_{n+1})=\la\ln\bigl(\b_n+s_{n+1}^2\bigr)
     -\frac{2\a_n}{\sqrt{\b_n}}\arctan\left(\frac{s_{n+1}}{\sqrt{\b_n}}\right),\label{def_phin}
\end{gather}
\end{subequations}
and $s_j(x,y;\la)$ are def\/ined implicitly through \Ref{eq:def-s}.

Equalities \Ref{XYPhi} can be verif\/ied by a direct, though tedious, computation.
Another, more elegant, way is to use the argument from \cite{Skl99}
based on imposing a set of constraints in an extended phase space,
see Appendix \ref{proof_canonicity}.

\subsection{Commutativity} The commutativity
$\Bcal_{\la_1}\circ\Bcal_{\la_2}=\Bcal_{\la_2}\circ\Bcal_{\la_1}$
of the BT follows from the preservation of the complete set of Hamiltonians
and the canonicity by the standard argument \cite{KS98,Skl99}
based on Veselov's theorem \cite{Ves91}
about the action-angle representation of integrable maps.

\subsection{Spectrality}\label{spectrality}
The spectrality property formulated f\/irst in \cite{KS98}
generalises the canonicity by allowing the parameter $\la$
of the BT to be a dynamical variable like $x$ and $y$.

Let us extend the symplectic space $V_{4n}$ from section \ref{Canonicity}
to a $(4n+2)$-dimensional space $V_{4n+2}$ by adding two more coordinates $\la$, $\mu$
and def\/ining the extension $\Om_{4n+2}$  of symplectic form~$\Om_{4n}$~\Ref{defOm4n}
as
\beq
    \Om_{4n+2}\equiv \Om_{4n}-\rd\mu\wedge\rd\la
    =\sum_{j=1}^n\bigl(\rd X_j\wedge\rd x_j-\rd Y_j\wedge\rd y_j\bigr)
    -\rd\mu\wedge\rd\la.
\label{bigform}
\eeq

Def\/ine the extended graph $\G_{2n+1}$ of the BT by equations \Ref{XTbcklnd}
and a new equation
\beq
   \mu=-\frac{\dd}{\dd\la}\Phi_\la(y;x).
\eeq

The 2-form $\Om_{4n+2}$ obviously vanishes on $\G_{2n+1}$, and
the manifold $\G_{2n+1}$ is lagrangian.

An amazing fact is that $\re^\mu$ is proportional to an eigenvalue of the matrix
$L(\la)$, see \Ref{eq:spectrality}.
In fact, the two eigenvalues of $L(\la)$ can be found explicitly
to be
\begin{subequations}\label{eigvL}
\begin{gather}
    \La=(\a_n^2+\b_n\la^2)\frac{1+\b_1 s_1^2}{\b_n+s_{n+1}^2}
    \prod_{j=1}^n \bigl(-s_j^{-2}e^{x_j+y_j}\bigr), \label{def_La}\\
    \tilde\La=(\a_1^2+\b_1\la^2)\frac{\b_n+s_{n+1}^2}{1+\b_1 s_1^2}
    \prod_{j=1}^n \bigl(-s_j^{2}e^{-x_j-y_j}\bigr),\label{def_Lat}
\end{gather}
\end{subequations}
see Appendix \ref{proof_spectrality} for the proof.

Having the explicit formulae \Ref{def_La} for $\La$ and \Ref{def_Phi}
for $\Phi_\la(y;x)$ one can easily verify that
\beq
 \La=(-1)^n(\a_n^2+\b_n\la^2)\,\re^\mu.
\eeq

\subsection{\BT\ as discrete time dynamics}
One of applications of a BT is that it might provide a discrete-time
approximation of a continuous-time integrable system
\cite{Suris2003,KPR2004}.
Indeed, iterations of the canonical map $\Bcal_\la$ generate a discrete time
dynamics. Furthermore, if we f\/ind a point $\la=\la_0$ that (a)
the map $\Bcal_{\la_0}$ becomes the identity map, and (b)
in a neighbourhood of $\la_0$ the inf\/initesimal map
$\Bcal_{\la_0+\eps}\sim\eps\{H,\cdot\}$ reproduces the Hamiltonian f\/low
with the Hamiltonian \Ref{hamiltonian}, we can claim that
$\Bcal_\la$ is a discrete time approximation of the BC-Toda lattice.
An attractive feature of this approximation is that,
unlike some others \cite{Suris2003}, the discrete-time system
and the continuous-time one share the same integrals of motion.

In our case $\la_0=\infty$. Letting $\eps=\la^{-1}$ and assuming the ansatz
\beq
    y_j=x_j+O(\eps), \qquad j=1,\ldots,n
\label{expand_yj}
\eeq
we obtain from \Ref{eq_sj} and \Ref{eq_s1} the expansion
\begin{subequations}\label{expand_s}
\begin{align}
    s_j&=\eps\re^{x_j}+O(\eps^2), \qquad j=1,\ldots,n \\
\intertext{and from \Ref{eq_sn+1} the expansion}
    s_{n+1}&=\eps(\a_n+\b_n\re^{-x_n})+O(\eps^2).
\end{align}
\end{subequations}

Substituting then expansions \Ref{expand_yj}
into equation \Ref{Sj1expand} we obtain
\begin{subequations}
\begin{align}
    S_j&=2\re^{-x_{j-1}}+O(\eps), \qquad j=2,\ldots,n+1 \\
\intertext{and substituting expansion \Ref{expand_s} for $s_1$
into formula \Ref{expandS1} for $S_1$ we obtain}
    S_1&=(2\a_1+\b_1\re^{x_1})+O(\eps).
\end{align}
\end{subequations}

Then from \Ref{def_Mj} we have
\beq
   -\eps M_j=\boldsymbol{1}+\eps\bigl(u\boldsymbol{1}+2A_j\bigr)+O(\eps^2), \qquad
   j=1,\ldots,n+1,
\eeq
where $A_j$ coincides with the matrix (given by \Ref{defAj} and \Ref{defA1n+1})
which describes the continuous-time
dynamics of the Lax matrix. From \Ref{Mel} we obtain then
\beq
   \ell(u;Y_j,y_j)=\ell(u;X_j,x_j)
   -2\eps\bigl(A_{j+1}\ell(u;X_j,x_j)-\ell(u;X_j,x_j) A_j\bigr)
     +O(\eps^2),
\eeq
for $j=1,\ldots,n+1$.
Comparing the result to \Ref{elldot} we get the expansion \Ref{expand_B}.

\section{Dual Lax matrix}\label{dualL}

Many integrable systems possess a pair of Lax matrices sharing the
same spectral curve with the parameters
$u$ and $v$ swapped like in \Ref{dual_curve},
see \cite{AHH90} for a list of examples and a discussion.
In particular, the periodic $n$-particle Toda lattice has two
Lax matrices: the `small' one,
of order~2~\cite{FTbook}, and the `big' one, of order $n$ \cite{Moe76}.
For various degenerate cases of the BC-Toda lattice `big' Lax matrices are
also known \cite{AvM82,RS94,RS2003,Moe76}.

In this section we present a new Lax matrix of order $2n+2$ for the most
general, 4-parametric BC-Toda lattice.
Here we describe the result, removing the
detailed derivation to Appendix \ref{dual_derivation}.

Let $E_{jk}$ be the square matrix of order $2n+2$ with the only nonzero
entry $(E_{jk})_{jk}=1$.
The Lax matrix $\Lcal(v)$£ is then described for the generic case $n\geq3$ as
\begin{gather}
    \Lcal(v)= \sum_{j,k=1}^n \Lcal_{jk}E_{jk} \notag\\
\phantom{\Lcal(v)}{}  = \sum_{j=2}^n \re^{x_j-x_{j-1}}E_{j,j-1}
      +\sum_{j=1}^n \bigl(-X_jE_{jj}+E_{j,j+1}\bigr) \notag\\
\phantom{\Lcal(v)=}{}      -\sum_{j=1}^{n -1}\re^{x_{j+1}-x_j}E_{2n+2-j,2n+1-j}
      +\sum_{j=1}^n \bigl(X_jE_{2n+2-j,2n+2-j}-E_{2n+2-j,2n+3-j}\bigr)
      \notag\\
\phantom{\Lcal(v)=}{}+\left(\a_n\re^{-x_n}+\frac{\b_n}{2}\re^{-2x_n}\right)
      \bigl(E_{n+1,n}-E_{n+2,n+1}\bigr) \notag\\
\phantom{\Lcal(v)=}{}+\frac{\b_n}{2}\re^{-x_n-x_{n-1}}
      \bigl(E_{n+3,n}-E_{n+2,n-1}\bigr)-E_{n+1,n+2} \notag\\
\phantom{\Lcal(v)=}{}-\left(\a_1\re^{x_1}+\frac{\b_1}{2}\re^{2x_1}\right)
      \bigl(E_{2n+2,2n+1}+v^{-1}E_{1,2n+2}\bigr)\notag\\
\phantom{\Lcal(v)=}{}+\frac{\b_1}{2v}\re^{x_1+x_2}
      \bigl(E_{2,2n+1}-E_{1,2n}\bigr)-vE_{2n+2,1}
\label{Lcal_expand}
\end{gather}
and consists of a bulk `Jacobian' strip (the main diagonal and two adjacent diagonals)
which reproduces the Lax matrix for the open Toda lattice together with boundary
blocks containing parameters $\a_1\b_1\a_n\b_n$.
We do not consider here the special case of
small dimensions $n=1,2$ when the two boundary blocks interfere with each other and
the structure of the Lax matrices becomes more complicated

To help visualise the matrix $\Lcal(v)$ we present an illustration for the case $n=3$,
using the shorthand notation $\xi_j\equiv\re^{x_j}$, $\eta_j\equiv\re^{y_j}$:
\begin{gather}
    \Lcal(v)=
    \begin{pmatrix}
         -X_1 & 1 & 0 & 0 & 0 &
           \scriptstyle-\frac{\b_1}{2v}\xi_1\xi_2 & 0 &
           \scriptstyle \frac{\a_1}{v}\xi_1+\frac{\b_1}{2v}\xi_1^2\\
         \scriptstyle\frac{\xi_2}{\xi_1} & -X_2 & 1 & 0 & 0 & 0 &
           \scriptstyle\frac{\b_1}{2v}\xi_1\xi_2 & 0\\
        0 & \scriptstyle\frac{\xi_3}{\xi_2} & -X_3 & 1 & 0 & 0 & 0 & 0\\
      0 & 0 & \scriptstyle \frac{\a_3}{\xi_3}+\frac{\b_3}{2\xi_3^2} &
      0 & -1 & 0 & 0 & 0 \\
      0 & \scriptstyle-\frac{\b_3}{2\xi_2\xi_3} & 0 &
      \scriptstyle -\frac{\a_3}{\xi_3}-\frac{\b_3}{2\xi_3^2} &
      X_3 & -1 & 0 & 0 \\
      0 & 0 & \scriptstyle\frac{\b_3}{2\xi_2\xi_3} & 0 &
      \scriptstyle-\frac{\xi_3}{\xi_2} & X_2 & -1 & 0 \\
      0 & 0 & 0 & 0 & 0 &
      \scriptstyle-\frac{\xi_2}{\xi_1} & X_1 & -1  \\
      -v &  0 & 0 & 0 & 0 & 0 &
      \scriptstyle-\a_1\xi_1-\frac{\b_1}{2}\xi_1^2 & 0
    \end{pmatrix}.
\end{gather}

The matrix $\Lcal(v)$ possesses the symmetry
\beq
    \Lcal(v)=-C_v\Lcal^t(v)C^{-1}_v,
\label{Lsym}
\eeq
where
\begin{gather}
    C_v=-vE_{2n+2,2n+2}+\sum_{j=1}^{2n+1}E_{j,2n+2-j}
       =\left(\begin{array}{ccccc|c}
         0 & 0 & \ldots & 0 & 1 & 0\\
         0 & 0 & \ldots & 1 & 0 & 0\\
         \ldots & \ldots & \ldots & \ldots & \ldots & \ldots \\
         0 & 1 & \ldots & 0 & 0 & 0 \\
         1 & 0 & \ldots & 0 & 0 & 0 \\
         \hline
         0 & 0 & \ldots & 0 & 0 & -v
    \end{array}\right)
\end{gather}
(note that $C_v^{-1}=C_{v^{-1}}$).

The matrix $\Lcal(v)$ shares the same spectral curve with the `small'
Lax operator $L(u)$ satisfying the determinantal identity \Ref{dual_curve}
and thus generates the same commuting Hamiltonians $H_j$.

The Lax matrix $\Lcal(v)$ of order $2n+2$ seems to be new.
When one or more of the constants $\a_1\b_1\a_n\b_n$ vanish
it degenerates (with a drop of dimension) into known Lax matrices
for simple af\/f\/ine Lie algebras \cite{AvM82,RS94,RS2003,Moe76}.
For the general 4-parametric
case a Lie-algebraic interpretation of $\Lcal(v)$ is still unknown.
In particular, it is an interesting question whether $\Lcal(v)$
satisf\/ies a kind of $r$-matrices Poisson algebra.

Inozemtsev \cite{Ino} presented a dif\/ferent Lax matrix for the BC-Toda
lattice, of order $2n$ instead of $2n+2$ and with a more complicated dependence on the spectral
parameter. The relation of these two Lax matrices is yet to be investigated.

For the dynamics \Ref{eq:dynamics-x}, \Ref{eq:dynamics-X} we have an analog of the
Lax equation \Ref{Ldot}:
\beq
    \dot{\Lcal(v)}=[\Acal(v),\Lcal(v)]
\eeq
with $\Acal(v)$ def\/ined as
\begin{gather}
    \Acal(v)=\sum_{j=1}^n\bigl(X_jE_{jj}-E_{j,j+1}
             -X_jE_{2n+2-j,2n+2-j}+E_{2n+2-j,2n+3-j}\bigr)
          +E_{n+1,n+1} \notag\\
\phantom{\Acal(v)=}{} +vE_{2n+2,1}-\frac{\b_1}{2}\re^{2x_1}\bigl(E_{2n+2,2n+1}+v^{-1}E_{1,2n+2}\bigr)
          +\frac{\b_n}{2}\re^{-2x_n}\bigl(E_{n+1,n}-E_{n+2,n}\bigr)\!\!\!
\end{gather}
and satisfying
\beq
    \Acal(v) C_v+C_v\Acal^t(v)=0.
\eeq

The analog of the formula \Ref{ML}
for the \BT\ is
\begin{subequations}\label{MLLM}
\begin{gather}
    \Mcal(v,\la)\Lcal(v;Y,y)=\Lcal(v;X,x)\Mcal(v,\la), \\
    \tilde\Mcal(v,\la)\Lcal(v;X,x)=\Lcal(v;Y,y)\tilde\Mcal(v,\la),
\end{gather}
\end{subequations}
where $\Mcal(v)$ is given by
\begin{gather}
    \Mcal(v)= \sum_{j,k=1}^n \Mcal_{jk}E_{jk}
     = -\sum_{j=2}^n \frac{\xi_j}{\eta_{j-1}}E_{j,j-1}
      +\sum_{j=1}^n \left(\frac{s_{j+1}}{\eta_j}-\frac{\xi_j}{s_j}\right)E_{jj}
      +E_{j,j+1} \label{Mcal_expand}\\
\phantom{\Mcal(v)=}{} +\sum_{j=1}^{n-1}\frac{\eta_{j+1}}{\xi_j} E_{2n+2-j,2n+1-j}
      +\sum_{j=1}^n \left(\frac{s_{j+1}}{\xi_j}-\frac{\eta_j}{s_j}\right)E_{2n+2-j,2n+2-j}
      -E_{2n+2-j,2n+3-j}  \notag\\
\phantom{\Mcal(v)=}{}      +\left(\frac{\a_n}{\xi_n}+\frac{\b_n}{2\xi_n^2}\right)
      \bigl(E_{n+1,n}-E_{n+2,n+1}\bigr)
      +\frac{\b_n}{2\xi_n\xi_{n-1}}
      \bigl(E_{n+3,n}-E_{n+2,n-1}\bigr)-E_{n+1,n+2} \notag\\
\phantom{\Mcal(v)=}{}      -\left(\a_1\xi_1{+}\frac{\b_1\xi_1^2}{2}\right)\!
      \bigl(E_{2n+2,2n+1}{+}v^{-1}E_{1,2n+2}\bigr)
      +\frac{\b_1\xi_1\xi_2}{2v}
      \bigl(E_{2,2n+1}{-}E_{1,2n}\bigr)-vE_{2n+2,1},\notag
\end{gather}
(using again the notation $\xi_j\equiv\re^{x_j}$, $\eta_j\equiv\re^{y_j}$)
and $\tilde \Mcal(v)$ is def\/ined as
\beq
    \tilde\Mcal(v)\equiv C_v\Mcal^t(v) C^{-1}_v.
\eeq

One of common ways to obtain a \BT\ is from factorising a Lax matrix in two dif\/ferent ways,
see \cite{AM97} for Toda lattices and \cite{Ves91} for other integrable models.
For our model we also have a remarkable factorisation, only instead of $\Lcal(v)$
we have to take its square:
\begin{subequations}\label{factorL}
\begin{gather}
    \la^2-\Lcal^2(v;X,x)=\Mcal(v,\la)\tilde\Mcal(v,\la),\\
    \la^2-\Lcal^2(v;Y,y)=\tilde\Mcal(v,\la)\Mcal(v,\la).
\end{gather}
\end{subequations}

\section{Discussion}\label{discussion}

The method for constructing a \BT\ presented in this paper seems to be quite
general and applicable as well to other integrable $sl(2)$-type chains with the boundary
conditions treatable within the framework developed in \cite{Skl87,Skl88}.

There is little doubt that a similar BT can be constructed
for the $D$-type Toda lattice and a~more general Inozemtsev's Toda lattice
\cite{Ino} with the boundary terms like
\be
   \frac{a_1}{\sinh^2\frac{x_1}{2}}+\frac{b_1}{\sinh^2 x_1}
   +\frac{a_n}{\sinh^2\frac{x_n}{2}}+\frac{b_n}{\sinh^2 x_n}
\ee
since those, as shown in \cite{KuzD}, can also be described in the formalism based on the
boundary $K$ matrices \Ref{defL} and the Poisson algebra \Ref{TT}.

The `big' Lax matrix $\Lcal(v)$ still awaits a proper
Lie-algebraic interpretation. Obtaining a BT from the factorisation
of $\la^2-\Lcal^2$ like in \Ref{factorL} might prove to be useful
for other integrable systems related to classical Lie algebras.

It is well known that the quantum analog of a BT is the so-called Q-operator
\cite{PG1992}, see also \cite{KS98}.
Examples of $Q$-operators for quantum integrable chains with a boundary
have been constructed recently for the XXX magnet \cite{DM05}
and for the Toda lattices of B, C and D types \cite{GLO2007}.
Our results for the BC-Toda lattice agree with those of
\cite{GLO2007}, the generating function of the BT being a classical limit
of the kernel of the $Q$-operator. Hopefully, our results will help to construct the
$Q$-operator for the general 4-parametric quantum BC-Toda lattice.

\appendix
\section{Proof of canonicity}\label{proof_canonicity}

Here we adapt to the BC-Toda case
the argument from \cite{Skl99} developed originally for the periodic case.
The trick is to obtain the graph $\G_{2n}$ of the BT as a projection of another
manifold in a~bigger symplectic space,
the mentioned manifold being Lagrangian for trivial reason.

Consider the 8-dimensional symplectic space $W_8$ with coordinates $XxYySsTt$
and the symplectic form
\beq
    \om_8\equiv\rd X\wedge \rd x+\rd S\wedge \rd s
           -\rd Y \wedge \rd y-\rd T\wedge\rd t.
\eeq

The matrix relation
\beq
  M(u,\la;T,t)\ell(u;Y,y)=\ell(u;X,x)M(u,\la;S,s)
\label{eq:M-ell_intertwine}
\eeq
is equivalent to 4 relations
\begin{subequations}\label{XSYT}
\begin{gather}
  X=-\la+s^{-1}\re^x+t\re^{-x}, \\
  Y=\la-s^{-1}\re^y-t\re^{-y}, \\
  S=2\la s^{-1}-s^{-2}\re^x-s^{-2}\re^y, \\
  T=\re^{-x}+\re^{-y},
\end{gather}
\end{subequations}
def\/ining a 4-dimensional submanifold $\Gcal_4\subset W_8$.
The fact that $\Gcal_4$ is Lagrangian, that is $\left.\om_8\right|_{\Gcal_4}=0$,
is proved by presenting explicitly the generating function
\beq
  f_\la(y,t;x,s)=\la(2\ln s-x-y)+s^{-1}(\re^x+\re^y)-t(\re^{-x}+\re^{-y}),
\eeq
such that
\beq
    X=\frac{\dd f_\la}{\dd x}, \qquad
    S=\frac{\dd f_\la}{\dd s}, \qquad
    Y=-\frac{\dd f_\la}{\dd y}, \qquad
    T=-\frac{\dd f_\la}{\dd t}.
\eeq

An alternative proof \cite{Skl99}
is based on the fact that $\ell(u)$ and $M(u,\la)$
are symplectic leaves of the same Poisson algebra \Ref{rll}.

Relation \Ref{eq:M-ell_intertwine} def\/ines thus a canonical
transformation from $XxSs$ to $YyTt$.

Let us take $n$ copies $W_8^{(j)}$ of $W_8$
decorating the variables $XxYySsTt$ with the indices $j=1,\ldots,n$
and impose on them $n$ matrix relations obtained from
\Ref{eq:M-ell_intertwine} by adding subscript $j$ to all variables.
We obtain then a Lagrangian manifold $\Gcal_{4n}=\otimes_{j=1}^n \Gcal_4^{(j)}$
in the $8n$-dimensional symplectic space $W_{8n}=\oplus_{j=1}^n W_8^{(j)}$
with the symplectic form $\om_{8n}=\sum\limits_{j=1}^n\om_8^{(j)}$
and the corresponding canonical transformation
with the generating function $\sum\limits_{j=1}^n f_\la(y_j,t_j;x_j,s_j)$.

Let us also introduce 4 additional variables
$T_0,t_0$ and $S_{n+1},s_{n+1}$ serving as coordinates in the
4-dimensional symplectic space $W_4$
with the symplectic form
$\om_4\equiv\rd S_{n+1}\wedge \rd s_{n+1}-\rd T_0\wedge \rd t_0$.
The relations
\beq
   T_0=\frac{2(\a_1+\b_1\la t_0)}{1+\b_1 t_0^2}, \qquad
   S_{n+1}=\frac{2(\la s_{n+1}-\a_n)}{\b_n+s_{n+1}^2}
\label{T0Sn}
\eeq
def\/ine then a 2-dimensional Lagrangian submanifold $\Gcal_2\subset W_4$
characterised by the generating function
$\phi=\phi_\la^{(0)}(t_0)+\phi_\la^{(n+1)}(s_{n+1})$
with $\phi_\la^{(0)}$ and $\phi_\la^{(n+1)}$ def\/ined by \Ref{def_phi0} and
\Ref{def_phin}, respectively:
\beq
    T_0=-\frac{\dd\phi_\la}{\dd t_0}, \qquad
    S_{n+1}=\frac{\dd\phi_\la}{\dd s_{n+1}}.
\eeq

We end up with
the $(8n+4)$-dimensional symplectic space $W_{8n+4}=W_{8n}+W_4$,
symplectic form $\om_{8n+4}=\om_{8n}+\om_4$, and
the $(4n+2)$-dimensional Lagrangian submanifold
$\Gcal_{4n+2}=\Gcal_{4n}\times\Gcal_2\subset W_{8n+4}$
def\/ined by the generating function
\beq
  F_\la=
  \phi_\la^{(0)}(t_0)+\phi_\la^{(n+1)}(s_{n+1})
   +\sum_{j=1}^n f_\la(y_j,t_j;x_j,s_j).
\label{big_genfun}
\eeq

The f\/inal step is to impose $2n+2$ constraints
\beq
   T_j=S_{j+1}, \quad t_j=s_{j+1}, \qquad j=0,\ldots,n,
\label{eq:constr-ts}
\eeq
which def\/ine a subspace $W_{6n+2}\subset W_{8n+4}$
of dimension $(8n+4)-(2n+2)=6n+2$
and respective $2n$-dimensional submanifold
$\Gcal_{2n}=\Gcal_{4n+2}\cap W_{6n+2}$.

Constraints \Ref{eq:constr-ts} allow to eliminate the variables $Tt$.
The space $W_{6n+2}$ splits then into the direct sum
$W_{6n+2}=V_{4n}+W_{2n+2}$ of
the space $W_{4n}$ with coordinates $X_jx_jY_jy_j$ ($j=1,\ldots,n$)
and $W_{2n+2}$ with coordinates $S_js_j$ ($j=1,\ldots,n+1$).
Using \Ref{eq:constr-ts} we obtain that
$\rd T_j\wedge \rd t_j-\rd S_{j+1}\wedge \rd s_{j+1}=0$
and therefore the symplectic form $\om_{8n+4}$ restricted on $W_{6n+2}$
\beq
    \left.\om_{8n+4}\right|_{W_{6n+2}}=\sum_{j=1}^n
    \bigl(\rd X_j\wedge \rd x_j-\rd Y_j\wedge y_j\bigr),
\eeq
degenerates: it vanishes on $W_{2n+2}$ and
remains nondegenerate on $V_{4n}$. In fact,
on $V_{4n}$ the form $\om_{8n+4}$ coincides
with the standard symplectic form \Ref{defOm4n}.
\beq
   \left.\om_{8n+4}\right|_{V_{4n}}=\Om_{4n}.
\eeq

After the elimination of the variables $Tt$
from equations \Ref{XSYT} and \Ref{T0Sn},
the resulting set of equations
def\/ining the submanifold $\Gcal_{2n}=\Gcal_{4n+2}\cap W_{6n+2}\subset W_{6n+2}$
coincides with equations~\Ref{XYSS} and~\Ref{expandS1}
def\/ining the BT.

As we have seen in Section \ref{BTdef},
the variables $S_js_j$ can also be eliminated
leaving a $2n$ dimensional submanifold $\G_{2n}\subset V_{4n}$
coinciding with the graph of the BT discussed in Section \ref{Canonicity}.
By construction, $\G_{2n}$ is the projection of $\Gcal_{2n}$
from $W_{6n+2}$ onto $V_{4n}$ parallel to $W_{2n+2}$.
Furthermore, $\G_{2n}$ is Lagrangian since $\om_{8n+4}$
vanishes on $\Gcal_{4n+2}$, therefore on $\Gcal_{2n}=\Gcal_{4n+2}\cap W_{6n+2}$,
and therefore on $\G_{2n}$.
The canonicity of the BT is thus established geometrically,
without tedious calculations.

The same argument as in \cite{Skl99} shows that
the generating function $\Phi_\la$ of the Lagrangian submanifold $\G_{2n}$
is obtained by setting $t_j=s_{j+1}$ in \Ref{big_genfun},
which produces formula \Ref{def_Phi}.

\section{Proof of spectrality}\label{proof_spectrality}
Here we provide the proof of formulae \Ref{eigvL} for the eigenvalues of $L(\la)$.
For the proof we use an observation from \cite{Skl99}
and show that the eigenvectors of $L(\la)$ are given by null-vectors
of $M_1(\pm\la,\la)$.

After setting $u=-\la$ in \Ref{def_Mj} the matrix $M_j$ becomes a projector
\beq
  M_j(-\la,\la)=\begin{pmatrix} -2\la+s_jS_j & s_j^2S_j-2\la s_j \\
              S_j & s_jS_j
        \end{pmatrix}
        =\begin{pmatrix}
          -2\la+s_jS_j \\ S_j
         \end{pmatrix}
         \begin{matrix}
         (1\quad s_j) \\ \phantom{0}
         \end{matrix}
\eeq
with the null-vector
\beq
   \s_j\equiv\begin{pmatrix} -s_j  \\ 1\end{pmatrix}, \qquad
   M_j(-\la,\la)\s_j=0.
\label{nullM}
\eeq

Let us set $u=-\la$ in the matrix equality \Ref{ML} and apply it to the vector $\s_1$.
By \Ref{nullM}, the right-hand side gives $0$. Therefore, $L(-\la)\s_1$ should
be proportional to the same null-vector $\s_1$ of $M_j(-\la,\la)$, and
$\s_1$ is an eigenvector of $L(-\la)$.

To f\/ind the corresponding eigenvalue $\La$, use the factorised expression \Ref{defL}
of $L(-\la)$ and apply it to $\s_1$.
Using \Ref{defell} we obtain
\beq
    \ell(-\la;Y_j,y_j)\s_j=-s_j\re^{-y_j}\s_{j+1},
\eeq
hence
\beq
    T(u;Y,y)\s_1=\s_{n+1}\prod_{j=1}^n \bigl(-s_j\re^{-y_j}\bigr).
\eeq

From \Ref{antipode} and \Ref{def_Mj} we obtain
\beq
    M_j^a(u,\la)=\begin{pmatrix}
             -u-\la+s_jS_j & -S_j \\
              2\la s_j-s_j^2S_j & u-\la+s_jS_j
        \end{pmatrix},
\label{def_Maj}
\eeq
hence
\beq
    M_j^a(\la,\la)=
       \begin{pmatrix}
         -2\la+s_jS_j & -S_j \\
         (2\la s_j-s^2_j)S_j & s_j S_j
       \end{pmatrix}
       =\begin{pmatrix}
          -1 \\ s_j
         \end{pmatrix}
         \begin{matrix}
         (2\la-s_jS_j \quad S_j) \\ \phantom{0}
         \end{matrix},
\eeq
the corresponding null-vector being
\beq
    \tilde\s_j\equiv \begin{pmatrix}
      S_{j} \\ s_jS_j-2\la
    \end{pmatrix}, \qquad
    M^a_j\tilde\s_j=0.
\eeq

A direct calculation using \Ref{defell} and \Ref{Sj1expand}
yields
\beq
    \ell_j^t(\la;Y_j,y_j)\tilde\s_{j+1}=s_j\re^{-x_j}\tilde\s_j
\eeq
and, consequently,
\beq
    T^t(\la;Y,y)\tilde\s_{n+1}=\tilde\s_1\prod_{j=1}^n\bigl(s_j\re^{-x_j}\bigr).
\eeq

From \Ref{defKpm} we get, respectively, the identities
\beq
    K_+(-\la)\s_{n+1}=\frac12(\b_{n}+s_{n+1}^2)\tilde\s_{n+1}, \qquad
    K_-(-\la)\tilde\s_1=2\frac{\a_1^2+\b_1\la^2}{1+\b_1s_1^2}\,\tilde\s_1.
\eeq

Using the above formulae we are able to move $\s_1$ through all the factors
constituting $L(-\la)$ and obtain the equality
\beq
    L(-\la;Y,y)\s_1=\La\s_1,
\eeq
where $\La$ is given by \Ref{def_La}.
Note that $\La$ is an eigenvalue of $L(\la)$ as well since
$\La(\la)=\La(-\la)$.
The second eigenvalue $\tilde\La$ \Ref{def_Lat} of $L(\la)$
is obtained from
\beq
\La\tilde\La=\det L(\la)\equiv d(\la)
  =(\a_n^2+\b_n\la^2)(\a_1^2+\b_1\la^2),
\eeq
see \Ref{detL}.

\section{Derivation of the dual Lax matrix}\label{dual_derivation}

To construct the `big' Lax operator $\Lcal(v)$
from the `small' one $L(u)$ we use the technique developed
for the periodic the periodic Toda lattice \cite{Skl99,KSS00}, with the
necessary corrections to accommodate the boundary conditions.

Let $\theta_1$ be an eigenvector of $L(u)$ with the eigenvalue $v$:
\beq
    L(u)\theta_1=v\theta_1, \qquad \theta_1=\binom{\phi_1}{\psi_1}.
\label{Ltheta1}
\eeq

Reading off the factors constituting the product $L(u)$,
see \Ref{defL}, \Ref{defT},
def\/ine recursively the vectors $\theta_j$
\beq
    \theta_j=\binom{\phi_j}{\psi_j}, \qquad
    j=1,\ldots2n+2,
\eeq
by the relations
\begin{subequations}
\begin{gather}
    \theta_{j+1}=\ell(u;X_j,x_j)\theta_j, \qquad
        j=1,\ldots,n,\\
    \theta_{n+2}=K_+(u)\theta_{n+1}, \\
    \theta_{n+j+3}=\ell^t(-u;X_{n-j},x_{n-j})\theta_{n+j+2}, \qquad
        j=0,\ldots,n-1,\\
\intertext{and close the circuit with the equation}
    v\theta_1=K_-(u)\theta_{2n+2},
\end{gather}
\end{subequations}
which is equivalent to \Ref{Ltheta1}.

A recursive elimination of $\psi_j$ results in the equations
\begin{subequations}
\begin{gather}
    u\phi_1=\phi_2-X_1\phi_1
           +\left(\frac{\a_1}{v}\re^{x_1}+\frac{\b_1}{v}\re^{2x_1}\right)\phi_{2n+2}
-\frac{\b_1}{v}\re^{2x_1}X_1\phi_{2n+1}
           +\frac{\b_1}{v}\re^{x_1+x_2}\phi_{2n}, \label{uphi_f}\\
    u\phi_j=\phi_{j+1}-X_j\phi_j+\re^{x_j-x_{j-1}}\phi_{n-1},\qquad
     j=2,\ldots,n \\
    u\phi_{n+1}=\phi_{n+2}+\a_n\re^{-x_n}\phi_n, \\
    u\phi_{n+2}=-\phi_{n+3}+X_n\phi_{n+2}
               +(\a_n\re^{-x_n}+\b_n\re^{-2x_n})\phi_{n+1} \notag\\
\phantom{u\phi_{n+2}=}{} -\b_n\re^{-2x_n}X_n\phi_n+\b_n\re^{-x_n-x_{n-1}}\phi_{n-1},
               \label{uphi_c}\\
    u\phi_j=-\phi_{j+1}+X_{2n+2-j}\phi_j-\re^{x_{j-3}-x_{j-4}}\phi_{j-1}, \qquad
        j=n+3,\ldots,2n+1, \\
    u\phi_{2n+2}=v\phi_1-\a_1\re^{x_1}\phi_{2n+1}.
\end{gather}
\end{subequations}

In order to simplify the 6-terms relations \Ref{uphi_f} and \Ref{uphi_c}
and make the matrix $\Lcal(v)$ more symmetric
we perform an additional reversible change of variables
\begin{subequations}
\begin{gather}
   \phi_1=\tphi_1+\frac{\b_1}{2v}\re^{2x_1}\tphi_{2n+1}, \\
   \phi_j=\tphi_j, \qquad j=2,\ldots,n+1, \\
   \phi_{n+2}=\tphi_{n+2}+\frac{\b_n}{2}\re^{-2x_n}\tphi_n, \\
   \phi_j=-\tphi_j, \qquad j=n+3,\ldots,2n+2.
\end{gather}
\end{subequations}

The resulting equations for $\tphi_j$ read
\begin{subequations}\label{eq_tphi}
\begin{gather}
    u\tphi_1=\tphi_2-X_1\tphi_1
           -\left(\frac{\a_1}{v}\re^{x_1}+\frac{\b_1}{2v}\re^{2x_1}\right)\tphi_{2n+2}
           -\frac{\b_1}{2v}\re^{x_1+x_2}\tphi_{2n}, \\
    u\tphi_2=\tphi_3-X_2\tphi_2+\re^{x_2-x_1}\tphi_1
           +\frac{\b_1}{2v}\re^{2x_1}\tphi_{2n+1}, \\
    u\tphi_j=\tphi_{j+1}-X_j\tphi_j+\re^{x_j-x_{j-1}}\tphi_{n-1}, \qquad
     j=3,\ldots,n \\
    u\tphi_{n+1}=-\tphi_{n+2}+\left(\a_n\re^{-x_n}+\frac{\b_n}{2}\re^{-2x_n}\right)\tphi_n, \\
    u\tphi_{n+2}=-\tphi_{n+3}+X_n\tphi_{n+2}
               -\left(\a_n\re^{-x_n}+\frac{\b_n}{2}\re^{-2x_n}\right)\tphi_{n+1}
              -\frac{\b_n}{2}\re^{-x_n-x_{n-1}}\tphi_{n-1},\\
    u\tphi_{n+3}=-\tphi_{n+4}+X_{n-1}\tphi_{n+3}
                -\re^{x_n-x_{n-1}}\tphi_{n+2}
                +\frac{\b_n}{2}\re^{-2x_n}, \\
    u\tphi_j=-\tphi_{j+1}+X_{2n+2-j}\tphi_j-\re^{x_{j-3}-x_{j-4}}\tphi_{j-1}, \qquad
        j=n+4,\ldots,2n+1, \\
    u\tphi_{2n+2}=-v\tphi_1-\left(\a_1\re^{x_1}+\frac{\b_1}{2}\re^{2x_1}\right)\tphi_{2n+1}.
\end{gather}
\end{subequations}

Introducing the vector $\Theta$ with $2n+2$ components
$\tphi_j$, $j=1,\ldots,2n+2$ we can rewrite relations
\Ref{eq_tphi} in the matrix form{\samepage
\beq
    \Lcal(v)\Theta=u\Lcal(v)\Theta
\label{eigvLcal}
\eeq
with the matrix $\Lcal(v)$ given by \Ref{Lcal_expand}.
It follows from \Ref{eigvLcal} that $u$ is an eigenvalue of $\Lcal(v)$.}

The rest of the formulae of Section \ref{dualL}
are obtained by a straitforward calculation
not much dif\/ferent from the periodic case \cite{Skl99,KSS00}.

\subsection*{Acknowledgements}

This work has been partially supported by the European Community (or European Union) through the FP6
Marie Curie RTN {\sl ENIGMA} (Contract number MRTN-CT-2004-5652).

\pdfbookmark[1]{References}{ref}

\LastPageEnding
\end{document}